\documentclass[aps,prl,showpacs,twocolumn,superscriptaddress,showpacs,groupedaddress,nofootinbib]{revtex4-1}  
\usepackage{graphicx}  
\usepackage{dcolumn}   
\usepackage{bm}        
\usepackage{amssymb}   
\usepackage{graphicx,amssymb,amsfonts,amsmath,amssymb,amscd,amstext}
\usepackage[colorlinks=true,citecolor=magenta, linkcolor=blue]{hyperref}

\pdfoutput=1
\def \be {\begin{equation}}
\def \ee {\end{equation}}
\def \bea {\begin{eqnarray}}
\def \eea {\end{eqnarray}}
\def \nn {\nonumber}
\def \la {\langle}
\def \ra {\rangle}
\def \del {\partial}

\def \b {\beta}

\def \eps {\epsilon}
\def \m {\mu}
\def \n {\nu}

\def \t {\tau}
\def\frac#1#2{{#1\over #2}}

\def\tr{\operatorname{tr}}

\newcommand{\oD}{\mathring{{\rm \nabla}}}

\begin{document}

\begin{flushright}
YITP-SB-14-51 
\end{flushright}

\title{Central Charge and Entangled Gauge Fields}

\author{Kuo-Wei Huang}
\affiliation{Department of Physics and Astronomy, \\
C. N. Yang Institute for Theoretical Physics,\\
Stony Brook University, Stony Brook, NY 11794, USA}

\fontsize{10pt}{12.2pt}\selectfont
\begin{abstract}
Entanglement entropy of gauge fields is calculated using the partition function
in curved spacetime with a boundary. We derive a Gibbons-Hawking-like term from
a Becchi-Rouet-Stora-Tyutin (BRST) action and a Wald-entropy-like codimension-2 surface 
term is produced. It is further suggested that boundary degrees of freedom localized 
on the entanglement surface generated from the gauge redundancy could be used 
to resolve a subtle mismatch in a universal conformal anomaly-entanglement 
entropy relation. 

\end{abstract}
\maketitle
\noindent
{\it{\bf{1: Introduction}}}

Despite being perhaps the weirdest consequence from quantum mechanics \cite{EPR}, 
the concept of entanglement plays an important role in many areas of physics: it is a 
key ingredient in quantum information, an order parameter in the phase transition
in many-body systems \cite{Vidal:2002rm}, and a measure of renormalization group 
flow in quantum field theories \cite{Casini:2012ei}. The entanglement entropy is also 
suggested as the origin of the black hole entropy \cite{Bombelli:1986} \cite{Srednicki:1993im}. 
Decomposing the full Hilbert space into pieces ${\rm{A}}$ and its complement ${\rm{B}}$, 
\bea
\label{partition}
\cal{H}=\cal{H_{\rm{A}}}\otimes \cal{H_{\rm{B}}} \ ,
\eea the entanglement entropy (or von Neumann entropy) is defined by 
\bea
S_{\rm{EE}} \equiv - \tr \rho_{\rm{A}} \ln \rho_{\rm{A}} \ .
\label{EEdef}
\eea 
The reduced density matrix, 
\bea 
\rho_{\rm{A}} \equiv \tr_{\rm{B}} \rho \ ,
\eea 
is a partial trace of the full density matrix $\rho$ over the degrees of freedom in the region ${\rm{B}}$. 
We will be interested in the entanglement entropy of continuous quantum gauge-field theories.
For gauge-field theories where the observables are Wilson loops, the partition given in \eqref{partition} 
can be an issue since such a partition would cut some loops. 
The Hilbert space of gauge fields is defined modulo the gauge
transformation. The direct factorization as a product of two Hilbert
spaces of the subsystems could be troublesome.  
See \cite{Buividovich:2008gq} \cite{Donnelly:2008vx} \cite{Casini:2013rba} \cite{Casini:2014aia} 
for attempts to address this issue in lattice gauge theories. 

It is often difficult to compute the entanglement entropy directly, in 
particular for spacetime dimensions higher than two. 
There are several alternative ways to compute the entanglement entropy. 
One is the so-called replica (conical) method  \cite{Callan:1994py} \cite{Calabrese:2004eu}
(see also \cite{Solodukhin:2008dh}). 
In this method one calculates the partition function
on an n-fold cover of the background space-time where a conical singularity is introduced. 
In this paper we will avoid the conical singularity by adopting a new 
method introduced recently in \cite{Casini:2011kv} 
(see also \cite{Casini:2010kt}).  We focus on obtaining the 
universal contribution to the entanglement 
entropy in Minkowski spacetime with a spherical entangling surface 
with a radius $\rm{R}$.  The main observation in \cite{Casini:2011kv} is 
that the full causal development, $\cal D$, connected to
the spherical region with the radius $\rm{R}$ can be conformally mapped 
to a new geometry ${\cal H}= {\rm{S}}^1 \times {\rm {H}}^{d-1}$, 
where ${\rm {H}}^{d-1}$ is a $(d-1)$ dimensional hyperbolic plane and $ {\rm{S}}^1$ 
is a circle associated with periodic Euclidean time. 
(One can also consider a conformal mapping to the static patch of de Sitter space. 
In this paper we will mostly focus on the hyperbolic geometry.)
The vacuum correlators in the causal development of the region inside 
the spherical surface in $d$-dimensional flat spacetime $\cal D$ are mapped to thermal 
correlators on ${\cal H}$. Moreover, the modular flow on $\cal D$ is shown to correspond to 
the time translation on $\cal H$. The correlators on $\cal H$ are periodic in time under an imaginary shift by a $2\pi {\rm{R}}$.
The radius ${\rm{R}}$ of the circle ${\rm{S}}^1$ then defines the temperature  
\bea
T={1\over \beta}={1\over 2\pi {\rm{R}}} \ .
\eea
Therefore, by conformally mapping a vacuum state of a conformal field theory (CFT) 
onto a thermal state on the hyperbolic spacetime, the computation 
of the entanglement entropy across the sphere then can be calculated as 
the thermal entropy of the hyperbolic space via
\bea
\label{ee}
S_{\rm{EE}}=(1-\b \del_\b)\ln Z(\b) |_{{\cal H},\b=2\pi {\rm{R}}} \ .
\eea 
On the other hand, AdS/CFT correspondence \cite{Maldacena:1997re}  also provides a way 
to calculate the entanglement entropy \cite{Ryu:2006bv} \cite{Ryu:2006ef}.  
We focus on the field-theory calculation in this paper.

A minor nuisance of the entanglement entropy in continuum
quantum field theories is its UV cut-off dependence. However, 
despite that the coefficients of power law divergences depend on regularization schemes, the 
log-divergent term in the entanglement entropy (in even space-time dimensions) 
is scheme-independent hence becoming a universal result. 
Moreover, the log-divergent term in the entanglement entropy with a spherical 
entangling surface in d-dimensional flat spacetime is shown to be dictated by 
the central charge  \cite{Casini:2011kv} \cite{Ryu:2006ef}
\bea
\label{4aa}
S_{\text{EE,\text{ln}}}=(-1)^{{d\over2}-1} ~4{\rm {A}} \ln ({{\rm{R}}\over\delta}) \ ,
\eea 
where $\delta$ is the divergence cut-off. The type-{\rm {A}} central charge ``{\rm {A}}'' is
defined as the conformal anomaly coefficient in even space-time dimensions in
\bea
\label{4aab}
\la T^\m_\m\ra=\sum_{\rm{i}} {\rm {B}}_{d({\rm{i}})} I_{d({\rm{i}})}-2(-)^{d\over2} {\rm {A}} {\rm{E}}_d\ ,
\eea 
where ${\rm{E}}_d$ is the Euler density and $I_{(d){\rm{i}}}$ 
are the Weyl invariants that define the type-{\rm {B}} anomalies in d-dimensions. 
The type-{\rm {B}} central charges do not contribute in our discussion 
since the spacetime in consideration will be conformally flat. We are interested in 
the universal contribution to the entanglement entropy from the type-{\rm {A}} 
central charge. 

Notice that Eqs \eqref{4aa} and \eqref{4aab} use a scheme without introducing 
the so-called type-{\rm {D}}  trace anomaly.  
In ${\rm {D}} =4$, the type-{\rm {D}}  anomaly is given by
\be
\la T^{\m}_{~\m}\ra_{\rm{D}}= \gamma~ \Box { {R}} \ ,
\ee 
where $R$ is the Ricci scalar and $\gamma$ is the type-{\rm {D}}  central charge.
This is also the minimal scheme used recently 
in \cite{Herzog:2013ed} \cite{Huang:2013lhw}  to obtain the 
general stress tensors from conformal anomalies based on the 
method discussed in \cite{BC}. (See also a recent paper \cite{Astaneh:2014sma} 
for related discussion. In their footnote 2, it is suggested that the log-divergent term 
in the entanglement entropy for the 4{\rm {D}}  U(1) gauge fields might be related to 
this scheme dependent type-{\rm {D}}  trace anomaly. It would be interesting to 
see if the approach considered in this paper can be further identified as the type-{\rm {D}}  
anomaly contribution.)

For 4D free gauge fields with spin $s=1$, the result predicted by the formula \eqref{4aa} is given by
\bea
\label{gauge}
S^{(s=1)}_{\rm{EE} ,\text{ln}}= - {31\over 45} \ln ({{\rm{R}}\over\delta}) \ .
\eea
This result can be independently confirmed using the vector heat kernel on manifolds 
with a conical singularity  \cite{Fursaev:2012mp} \cite{DeNardo:1996kp}. 
However, to our knowledge, a field-theory calculation via the approach 
developed in \cite{Casini:2011kv} reproducing this result is absent (besides 
directly adopting the anomaly coefficients). In \cite{Eling:2013aqa}, a direct 
modification of the stress tensor is suggested to obtain this result.  
In this paper, we would like to see more closely 
what new ingredients are needed to give $\eqref{gauge}$ without introducing 
the subtle conical singularity.  Our main motivation is that finding a way to
improve an alternative method of calculating the entanglement entropy might 
shed light on defining the entanglement entropy  directly 
for general quantum gauge-field theories. 

We organize this paper as follows. In the next section, starting from the 
action principle, we revisit the formulation of gauge fields
in general curved spacetime with a boundary. We argue that the corresponding
Gibbons-Hawking-like term should be derived from the action instead of adding it by hand.
Although we do not adopt  in the paper the replica method 
that introduces the conical contact entropy \cite{Kabat:1995eq},
we discuss the similarity of the codimension-2 surface term in the action with the
contact term in Sec.3.  In Sec.4, we calculate the partition
function and thermal entropy on ${\rm S}^1\times {\rm{H}}^3$ using the heat kernel method.
The main result of this section is that the entropy \eqref{mis} has a mismatch
compared to the universal conformal anomaly prediction \eqref{gauge}. (We refer readers
to \cite{Hung:2014npa} for the corresponding calculation of the conformally
coupled scalar field and fermion on ${\rm S}^1\times {\rm{H}}^3$. The resulting entropy results
are consistent with the anomaly prediction \eqref{4aa}.) A resolution is suggested
in Sec.5, where it is argued that one
should include the edge-mode contributions due to the gauge symmetry. We also show that 
adding these edge modes allows us to reproduce the universal log-divergent term of the R\'enyi
entropy of gauge fields. We conclude this paper with some remarks on the entropy
mismatch in the static patch of de Sitter space found by \cite{Dowker:2010bu}, the 
issue of the Hilbert space decomposition, and black hole entropy.
\\

\noindent
{\it{\bf{2: Gauge Fields in Curved Spacetime with a Boundary Revisited}}}

Our starting point is the standard action of the U(1) gauge field $A_\m$ on a general 
spacetime background $\cal M$,
\bea
S={1\over 4} \int_{\cal {M}} F_{\m\n}F^{\m\n} \ ,
\eea 
where $F_{\m\n}=[\nabla_\m, \nabla_\n ]$ is the field strength.
Due to the gauge symmetry, $\delta A_\m=\partial_\m \lambda$ where $\lambda$ 
is the gauge parameter, we add the Lorenz gauge-fixing term given by
\bea
S_{\rm{gf}}={1\over2} \int_{\cal {M}} (\nabla_\m A^\m)^2 \ .
\eea 
The gauge-fixing procedure introduces the standard Fadeev-Popov ghosts $\bar b$ 
and $b$ that are anti-commuting scalars. The full gauge-fixed action is
\bea
\label{act}
S = \int_{\cal M}  {1 \over 4} F_{\mu\nu} F^{\mu\nu} +\int_{\cal M} \left( {1 \over 2} \big(\nabla_\mu A^\mu\big)^2
+\partial^\mu \bar b \partial_\mu b\right) \ .
\eea 
The above action has the following BRST symmetry parametrized by 
an infinitesimal anticommuting constant parameter $\eps$:
\bea
\label{brst}
\delta_{\eps} A_\m=(\partial_\m b) \eps,~~\delta_{\eps} b=0,~~ \delta_{\eps} \bar b= -(\nabla^\m A_\m) \eps,
\eea 
provided that a boundary condition is imposed. Either  $\partial_{\rm{n}} {{b}}|_{\partial M}=0$ 
or $\nabla_\mu A^\mu|_{\partial M}=0$. (We denote $\nabla_{\rm{n}}= {\rm{n}}^\m\nabla_\mu$.) 
In fact, when one writes the bulk ghost action 
as $ - \int_{\cal M} \bar {b}\Box {{b}}$, integration by 
parts is used and we should also
impose a boundary condition:  Either $\partial_{\rm{n}} {{b}}|_{\partial M}=0$ or  $\bar {{b}}|_{\partial M}=0$, 
whose BRST symmetry requires $\nabla_\mu A^\mu|_{\partial M}=0$. We will adopt 
$\partial_{\rm{n}} {{b}}|_{\partial M}=0$ in the following discussion.

We next emphasize that, different from the action of the non-minimally coupled 
scalar fields, the original gauge-field action \eqref{act} does not have second derivatives of the metric, 
so  one does not really need to add by hand a Gibbons-Hawking-like term in the action. 
However, as we will discuss later on obtaining the partition function using the heat kernel method, it is most natural to 
use an action which involves a second-order differential operator given by
\bea
{\rm D}_{\m\n}\equiv \Box\delta_{\m\n}-{ {R}}_{\m\n}\ .
\eea 
(We have denoted $\Box=\nabla^\mu \nabla_\mu$ 
and ${ {R}}_{\m\n}$ is the Ricci curvature tensor generated by $[\nabla_\m,\nabla_\n] A^\n= - { {R}}_{\m\n} A^\n$.) 
This operator is produced from integrating the standard action \eqref{act} by parts. 
The action using the operator ${\rm D}_{\m\n}$ naturally needs a Gibbons-Hawking-like term. However, 
one should not add a new term during an immediate calculation. Our resolution is that the 
Gibbons-Hawking-like term should be $derived$ in the gauge-field case.  
More precisely, we consider
\bea
\label{gaa}
&&\int_{\cal M} \left( {1 \over 4} F_{\mu\nu} F^{\mu\nu} + {1 \over 2} \big(\nabla_\mu A^\mu\big)^2\right)= -{1 \over 2}  \int_{\cal M}  A^\mu {\rm D}_{\m\n}  A^\nu\nn\\
&&~~~ -  {1 \over 2} \int_{\partial M}  \rm{n}_\m \Big( A_\nu  \nabla^\nu A^\m-A^\m \nabla^\nu A_\nu-A_\nu \nabla^\mu A^\nu\Big) \nn\\
&=&-{1\over 2}\int_{\cal M} A^\mu {\rm D}_{\mu\nu} A^\nu+ {1\over 2} \int_{\partial M} \Big( \rm{K}_{{\rm{i}}{\rm{j}}} A^{\rm{i}} A^{\rm{j}}+\rm{K} A^2_n \Big)\nn\\
&&+\int_{\partial M}  \Big( A_n \oD^{\rm{i}} A_{\rm{i}} +{1\over 4} \partial_n A^2 \Big) -{1\over 2} \int_\Sigma  \rm{n}^\mu \tilde n^\nu A_\mu A_\nu\ ,
\eea 
where we have denoted $\oD_{\rm{i}}$ as the boundary covariant derivative. Indices ${\rm{i}}, {\rm{j}}, {\rm{k}}$ represent the tangential directions. Choosing to extend ${\rm{n}}_{\mu}$ in such a way that $\nabla_{\rm{n}} {\rm{n}}_\mu=0$, we can write the extrinsic curvature as $\rm{K}_{\mu\nu}=\nabla_{(\mu} n_{\nu)}$. 
In obtaining the last line we have performed integration by parts on the boundary and $\tilde {\rm{n}}^\nu$ represents a normal vector used to define a codimension-2 manifold. We obtain the Gibbons-Hawking-like term that provides the cancellation 
involving the normal derivative of the metric variation on the boundary. 
To our knowledge, no literature has mentioned this kind of treatment regarding 
a gauge-field action in curved spacetime with a boundary. 
(See \cite{Barvinsky:1995dp} and \cite{Jacobson:2013yqa} for related discussions.)

We consider the following boundary conditions:
\bea
\label{bc}
&& A_{{\rm{n}}}|_{\partial M}=0,~ \partial_{\rm{n}} b|_{\partial M}=\partial_{\rm{n}} \bar b|_{\partial M}=0, \nn\\
&& (\nabla_{\rm{n}} A_{\rm{i}}+ \rm{K}_{{\rm{i}}{\rm{j}}} A^{\rm{j}})|_{\partial M}=0 \ .~~~
\eea 
These boundary conditions are referred to as the ``absolute" boundary conditions in \cite{Vassilevich:2003xt}.
Notice another type of boundary conditions called the ``relative" boundary conditions \cite{Vassilevich:2003xt}: $A_{{\rm{i}}}|_{\partial M}=0,~b|_{\partial M}=\bar b|_{\partial M}=0,~ (\partial_{\rm{n}} A_{\rm{n}}+ \rm{K} A_{\rm{n}})|_{\partial M}=0 $. However, for the entanglement entropy calculation, here we do not want to impose such strong constraints on the entangling surface.
\\

\noindent
{\it{\bf{3: Conical Contact Entropy}}}

The main reason we consider  in this paper a different approach to 
study the entanglement entropy of gauge fields instead of 
adopting the traditional replica method is that one 
is already able to obtain the expected central charge-entropy relation 
$\eqref {gauge}$ using the conical method \cite{Fursaev:2012mp} \cite{DeNardo:1996kp}. 
On the other hand, it is well known that the conical singularity 
causes subtle issues such as generating a contact 
term \cite{Kabat:1995eq}. Notice that a contact term also appears in the 
case of the non-minimally coupled scalar field \cite{Solodukhin:2011gn}.  
However, as we will discuss later, there is no entropy mismatch regarding 
the expected central charge-entropy relation in the conformal scalar field's 
case if using the hyperbolic space approach.  
In short, here we would like to resolve the entropy mismatch of 
gauge field's entanglement entropy without touching the contact term issue.

Although we will not focus on the conical approach, 
we would like to make a short remark on
the similarity of the codimension-2 
surface term in $\eqref{gaa}$ with the contact term of gauge fields.  
Let us briefly review the contact term contribution to the 
entanglement entropy for gauge fields \cite{Kabat:1995eq}. 
See also \cite{Donnelly:2012st} \cite {Kabat:2012ns} \cite{Lewkowycz:2013laa} \cite{Lee:2014zaa}  for recent related discussions.

If all surface terms are dropped, the $U(1)$ action is simply given by
\bea
\label{action again}
S_{\cal M}= -{1\over 2} \int_{\cal M}  A^\m {\rm{D}}_{\m\n} A^\n -\int_{\cal M} \bar b\Box b \ . 
\eea
On a manifold $\cal M$ with a conical singularity, we are 
interested in the heat kernel for first-order change 
of the conical angle $\beta$ away from $2\pi$. 
The conical deficit introduces a singular curvature at 
the tip of a cone. The curvature can be expanded as  \cite{Fursaev:1995ef}
\bea
\label{Rmn}
R_{\m\n}=\bar R_{\m\n}+ {(2\pi-\beta)} g^{\perp}_{\m\n}\delta_{\Sigma} + {\cal O}{(2\pi-\b)^2} \ ,
\eea  
where $\bar R_{\m\n}$ vanishes in flat spacetime.  $g^{\perp}_{\m\n}$ denotes a projection onto the directions perpendicular to the codimension-2 boundary.
The higher-order terms in $\eqref{Rmn}$ do not affect the entanglement entropy. 
The entropy formula in the conical method reads
$S_{\rm{cone}}=(1-\b \del_\b)\ln Z(\b) |_{\b=2\pi}$. The ghost fields do 
not contribute to the contact entropy. The partition fucntion of gauge 
fields can be written, using \eqref{Rmn}, as ($\bar {\rm{D}}_{\m\n}\equiv \Box\delta_{\m\n}-\bar { {R}}_{\m\n}$.)
\bea
\label{contact}
&&\int {\cal D}A \exp\Big[{1\over 2} \int_{\cal M}  A^\m (\bar {\rm{D}}_{\m\n}- {(2\pi-\beta)} g^{\perp}_{\m\n}\delta_{\Sigma}) A^\n \Big] \nn\\
&&=\bar Z_A-{(\pi-{\beta\over2})}\int_{\Sigma} \Big\la g^{\perp}_{\m\n} A^\m A^\n\Big\ra \ ,
\eea 
where the first term $\bar Z$ denotes the ``regular'' contribution. 
The second term leads to the contact entropy. ($\Sigma$ denotes a codimension-2 surface.) Let us also remark that this 
term is intimately related to the expectation value of Wald entropy \cite{Donnelly:2012st} \cite{Wald:1993nt},
\bea
\la S_{\text{Wald}}\ra= -2\pi \la\int_{\Sigma} {\partial {\cal L}\over \partial { {R}}_{\m\n\lambda\rho}} \eps_{\m\n} \eps_{\lambda\rho}\ra=-{\pi}\int_{\Sigma} \la g^{\perp}_{\m\n} A^\m A^\n\ra  \ . \nn\\
\eea  
We see that the contact term has the similar form as the codimension-2 term in the action \eqref{gaa}. 
However, in our approach without introducing a conical singularity, such a surface 
term will be killed by imposing boundary conditions. The 
treatments of the surface terms may be different depending on whether we are 
considering a physical boundary, an entanglement boundary or a conical singularity. 
We leave the problem of how the surface terms interact with 
the conical singularity as a future problem.
\\

\noindent
{\it{\bf{4: Gauge Fields on Hyperbolic Geometry and Entropy Discrepancy}}}

We are interested in the entanglement entropy of gauge fields on ${\cal R}^{1,3}$ with the entangling  
surface $S^2$ with radius R, at a time slice $t=0$.  
Using the approach developed in \cite{Casini:2011kv}, the 
computation of the entanglement entropy is mapped to calculating the 
thermal entropy on the hyperbolic space.
The original flat spacetime metric written in polar coordinates is given by
\bea
\label{flat}
{\rm{d}}s^2=-{\rm{d}}t^2+{\rm{d}}r^2+r^2 {\rm{d}}^2\Omega_{2} \ ,
\eea 
where ${\rm{d}}^2\Omega_{2}$ is the metric of the sphere with unit radius.
The transformations that map the geometry into the hyperbolic space are given by 
\bea
\label{mapping}
t&=& {\rm{R}} {\sinh ({\t\over R})\over \cosh \rm{u}+\cosh({\t\over R})}\ ,\\
r&=& {\rm{R}} {\sinh ({\rm{u}})\over \cosh \rm{u}+\cosh({\t\over R})} \ .
\eea 
The metric then becomes 
\bea
\label{hyper}
{\rm{d}}s^2=\Omega^2\Big(-{\rm{d}}\t^2+{\rm{R}}^2({\rm{d}}\rm{u}^2+\sinh^2\rm{u} ~{\rm{d}}^2\Omega_{2})\Big) \ .
\eea
The prefactor, $\Omega={(\cosh \rm{u}+\cosh{\t\over {\rm{R}}})^{-1}}$,
can be eliminated via the conformal transformation and the 
resulting metric is $\rm{S}^1 \times \rm{H}^3$. Notice the limits
\bea
\label{limit}
\t&=&\pm \infty \to (t=\pm {\rm{R}},r=0) \ , \\
\label{limitu}
\rm{u}&=&\infty \to (t=0, r={\rm{R}}) \ ,
\eea 
confirm that the full causal development $\cal D$ is indeed covered in $\cal H$ after the conformal mapping. 

We will use the heat kernel method on $\rm{S}^1 \times \rm{H}^3$ 
to obtain the partition function and the entropy.  Gravity will not be dynamical.
We denote the kernel as $K({\rm{x}},{\rm{y}};s)$ on a fixed spacetime 
background $\cal M$ satisfying the heat equation 
\bea
(\partial_s+{\text D})K({\rm{x}},{\rm{y}};s)=0 \ ,
\eea 
where $\rm{D}$ is a second-order differential kinetic operator. 
A boundary condition is imposed, $K({\rm{x}},{\rm{y}};0)=\delta({\rm{x}},{\rm{y}})$.
The trace of the heat kernel is given by 
\bea 
{\cal K}(s)\equiv \int_{\cal M} K({\rm{x}},{\rm{x}};s)=\sum_{\rm{i}} e^{-s\lambda_{\rm{i}}} \ ,
\eea
with summation over all eigenvalues $\lambda_{\rm{i}}$ of the 
operator $\rm{D}$ including possible degeneracy. 
Notice that the parameter $s$ must have dimensions of length squared 
if the argument of the exponential is to be dimensionless.
The partition function can be expressed via the heat kernel, 
\bea
\label{z}
\ln Z=- {1\over 2}\sum_{\rm{i}} \ln \lambda_{\rm{i}}={1\over 2} \int_{0}^{\infty} {ds\over s}{\cal K}{(s)} \ .
\eea

For the U(1) gauge fields on $S^1\times H^3$, after imposing the 
boundary conditions, the action reduces to \eqref{action again}.  
The partition function can be written as
\bea
Z=\rm{Det}(-\Box_s) \int {\cal D} A_\m \exp\Big[{{1\over2}\int_{\rm{S}^1 \times \rm{H}^3} A^\m D_{\m\n} A^\n}\Big] \ ,
\eea 
where the factor ${\rm {Det}}(-\Box_s)$ stands for the Faddeev-Popov determinant. 
Factoring out the temporal index and performing a Gaussian integral over $A_\t$ yields
\bea
Z
={\rm {Det}}(-\Box_s)^{1/2} \int{\cal D} A_{\rm{i}} \exp\Big[{{1\over2}\int_{\rm{S}^1 \times \rm{H}^3} A^{\rm{i}} {\rm D}_{\rm{i}\rm{j}} A^{\rm{j}}}\Big] \ .
\eea
Next we write
\bea
\label{Z}
\ln Z(\b) ={1\over 2}\Big(\ln {\rm {Det}}(-\Box_s)-\ln {\rm {Det}}(\rm{D}_{\rm{i}\rm{j}})\Big)={1\over 2} \int_{0}^{\infty} {ds\over s} {{\cal K}(s)} \ , \nn\\
\eea  
where we decompose the total heat kernel by
\bea
{\cal K}(s)= K_{\rm{i}\rm{j}} (S^1)K^{\rm{i}\rm{j}}(H^3)- K_{s} (S^1)K_{s}(H^3) \ ,
\eea  
with $K_{\rm{i}\rm{j}} (S^1)$ being a short hand for $\tr\int_{S^1} K_{\rm{i}\rm{j}}(\t,\t;s)$ 
and $K_{\rm{i}\rm{j}} ({\rm{H}}^3)$ for $\tr\int_{{\rm{H}}^3} K_{\rm{i}\rm{j}}(\rm{x},\rm{x};s)$.
The same notation applies on $K_s$ (scalar) parts. 
The volume simply factorizes in the heat kernels 
since the hyperbolic space is homogeneous.

The heat kernel on $S^1$ can be evaluated using the method of 
images preserving the periodic boundary condition. 
The result is given by an infinite sum on an infinite line shifted by ${2\pi R \rm{n}} ~(= \rm{n}\b)$,
\bea
\label{ij1}
K_{\rm{i}\rm{j}} (S^1)={2\b \over (4\pi s)^{1/2}} \sum_{\rm{n}=1}^{\infty} e^{- {\rm{n}^2\beta^2\over 4s}}=K_{s} (S^1) \ .
\eea
The $\rm{n}=0$ part is ignored because it will not contribute to the entanglement entropy. 

The heat kernels $K^{{\rm{i}}{\rm{j}}}({\rm{H}}^3)$ and $K_{s}({\rm{H}}^3)$ 
can be found in the literature \cite{CH1994}  \cite{GN} \cite{Giombi:2008vd} and are given by
\bea
\label{ij2}
K_{\rm{i}\rm{j}}({\rm{H}}^3)={{e^{-{s\over {\rm{R}}^2}}+2+4{s\over {\rm{R}}^2}}\over (4\pi s)^{3/2}}~;~
K_{s}({\rm{H}}^3)={{e^{-{s\over {\rm{R}}^2}}}\over (4\pi s)^{3/2}} \ . \nn\\
\eea
Plugging these results into $\eqref{Z}$ gives
\bea
\label{pati}
\ln Z(\b)
={{2\pi^2 {{\rm {R}}^2+15\b^2}\over 90 {\rm{R}}^2 \b^3}} \rm{Vol}({\rm{H}}^3) \ .
\eea 
We have to introduce an IR cut-off since the volume of $\rm{H}^3$ is divergent. We let \cite{Casini:2011kv}
\bea
\label{IR}
\cosh(\rm{u}_{\rm{max}})={{\rm {R}}\over \delta} \ .
\eea 
The scale of the hyperbolic curvature is set to be $\rm {R}$. 
We obtain a log term from 
\bea
\rm{Vol}({\rm{H}}^3)= - 2\pi  {\rm{R}}^3 \ln({{\rm{R}}\over \delta})+...
\eea
Finally, using \eqref{ee}, we obtain the entropy
\bea
\label{mis}
S_{\rm{EE},\text{ln}}=-{16\over 45} \ln({{\rm{R}}\over \delta})\ .
\eea  
We find a mismatch when we compare this result with the 
conformal anomaly prediction $\eqref{gauge}$.
\\

\noindent
{\it{\bf{5: Edge Entropy from Entangling Surface}}}

Here we suggest a way to resolve the mismatch found in $\eqref{mis}$. Our 
resolution is based on including edge modes localized on the entangling surface. 

Recall that the gauge symmetry results in the gauge fixing condition $\nabla^\m_{(\cal M)} A_\m=0$.
The gauge redundancy is determined by  
\bea
\Box_{(\cal M)} \lambda=0 \ ,
\eea 
where $\Box_{(\cal M)}$ is the D'Alembertian operator on ${\cal M}$. 
In the bulk, the residual gauge freedom is fixed by imposing a boundary condition on the 
boundary of ${\cal M}$, which is ${\cal {\partial M}}=\rm{S}^1 \times \rm{S}^2$. 
(We take $\rm{u}\to \infty$ as the boundary.) That is, we fix the residual gauge by 
imposing a constraint on the boundary, $\lambda_{\cal {\partial M}}=\bar \lambda$, 
where $\bar \lambda$ 
still satisfies $\Box_{(\cal{\partial M})} \bar \lambda=0$. 
Notice that by taking a large-$\rm{u}$ limit, the 4D metric $\eqref{hyper}$ (after eliminating the conformal factor) effectively 
reduces to 2D since the radius of the time circle is much smaller than the 
radius of $\rm{S}^2$, $\rm{R} \sinh \rm{u}_{\rm max}$. Therefore, in the sense of the Kaluza-Klein 
massive modes decoupling, the effective boundary becomes a 2-sphere and the gauge 
redundancy condition results in
\bea
\label{key}
\Delta_{(\rm{S}^2)} \bar \lambda=0 \ ,
\eea 
where $\Delta_{(\rm{S}^2)}$ is the Laplacian operator on $\rm{S}^2$.
Recall also that in the large-$\rm{u}$ limit, Eq \eqref{limitu} shows 
that it corresponds to the $\rm{t}=0$ slice that is used to define the 
original time-independent entanglement entropy with a static entangling surface. 

We interpret that the entangling boundary plays a role 
to encode these boundary redundancy modes.
In other words, the freedom of choosing different boundary data, $\bar\lambda$, in \eqref{key} is 
interpreted as having edge degrees of freedom on the entangling boundary. 
We suggest that these boundary modes give additional contributions to the 
entanglement entropy and can be used to resolve the mismatch $\eqref{mis}$.

The bulk partition function does not capture these boundary modes because the
surface action is set to zero using boundary conditions.
We treat \eqref{key} as a field equation on $S^2$
and define the corresponding partition function again by \eqref{z}. The question then can be
reduced to finding the corresponding eigenvalues using the heat kernel method.
 
The heat kernel on $\rm{S}^2$ is essentially given by solving the standard eigenvalue problem of the 
Laplacian on $\rm{S}^2$. The eigenvalues are ${l}({l}+1)$ with the orbital quantum number ${l}$ with the 
degeneracy given by $(2{l}+1)$. The eigenfunction is the familiar spherical harmonic. (See \cite{CR} for 
heat kernels in different spacetime manifolds.)  The heat kernel (density) that we need is given by 
\bea
\label{s2}
K({\rm{S}}^2)= {1\over{4\pi r^2}} \sum^{\infty}_{l=0} (2l+1) e^{-s{{l}({l}+1)\over r^2}} \ .
\eea 
We will be interested in the small-$s$ expansion. We use the Euler-MacLaurin formula
\bea
\label{em}
\sum^{\infty}_{{l}=0} f({l})=\int^{\infty}_{{l}=0} d{l} f({l})+{1\over2} f(0)-{1\over 12} f'(0)+...
\eea
with a function $f({l})$ satisfying $f^{(n)}(\infty)=0$ for arbitrary $\rm{n}$. 
We focus on the scheme-independent log divergence; the 
higher order terms in \eqref{em} are irrelevant. 
From \eqref{s2} and \eqref{em} we obtain
\bea
\label{ss2}
K({\rm{S}}^2)= \frac{12 r^4+4 r^2 s+s^2}{48 \pi r^4 s} +...
\eea 
Define the partition function on $S^2$ as
\bea
\ln Z({\rm{S}}^2)= {1\over 2} {\rm{Vol}}({\rm{S}}^2)\int^{\infty}_{\epsilon^2} {ds\over s} K(\rm{S}^2) \ ,
\eea 
where ${\rm{Vol}}(\rm{S}^2)=4\pi r^2$ is simply the area of the entangle surface. 
The s-independent term in $\eqref{ss2}$ gives the log divergence. 
Notice that the gauge parameter $\lambda$ is understood as  
the ghost $b$ so it contributes as a $negative$ massless scalar field on $S^2$.  
We obtain the log-divergent term from the edge modes,
$ \ln Z(\rm{S}^2)\to - {1\over 6} \ln({{\rm{R}}^2\over\epsilon^2})$. 
A dimensional scale R is inserted to have a dimensionless argument. 
We see that the log divergent term is independent of the radius of the entangling surface.
We identify the UV cut-off $\epsilon$ with the cut-off $\delta$ in 
regularizing ${\rm {Vol}}(\rm{H}^3)$, $\epsilon={\delta}$. The edge 
correction is given by
\bea
\label{13}
\Delta S^{(s=1)}_{\rm{EE},\text{ln}}= -{1\over 3} \ln\Big({{\rm{R}}\over \delta}\Big) \ ,
\eea 
which resolves the mismatch.
\\

Let us generalize our discussion to the R\'enyi entropy defined by
\bea
S_{\rm{q}}= {\ln \tr \rho_{\rm{A}}^{\rm{q}}\over 1-\rm{q}} \ .
\eea
It has the following simple relation to the entanglement entropy 
(assuming a satisfactory analytic continuation can be performed):
\bea 
S_{\rm{EE}}=\lim_{\rm{q}\to 1} S_{\rm{q}} \ .
\eea  
Having the hyperbolic partition function, we can calculate the R\'enyi entropy via 
\bea
S_{\rm{q}}= {\ln Z(\rm{q} \b)- \rm{q}\ln Z (\b)\over (1-\rm{q})} |_{\b\to 2\pi {\rm{R}}} \ .
\eea
Using \eqref{pati}, we obtain
\bea
S_{\rm{q}}= \frac{(\rm{q}+1) \left(31 \rm{q}^2+1\right)}{360 \pi \rm{q}^3 \rm{R}^3} {\rm {Vol}}({\rm{H}}^3) \ .
\eea 
Let us also include the edge modes. We should view the edge contribution 
as a universal contribution (the log-divergent term) in the sense that it is independent of $\beta$ or the radius of 
the entangling surface. It then should be also independent of the parameter $q$ 
inserted in temperature $T={1\over 2\pi {\rm R} q}$. By adding the edge contribution 
$\eqref{13}$, the full log-divergent part of the R\'enyi entropy becomes
\bea
S^{(s=1)}_{\rm{q},\text{ln}}=-{1+\rm{q}+31\rm{q}^2+91 \rm{q}^3\over {180 \rm{q}^3}}\ln ({{\rm{R}}\over \delta }) \ .
\eea 
This result (giving the coefficient -31/45 when taking q=1) is now 
consistent with the R\'enyi entropy result obtained in \cite{Fursaev:2012mp} \cite{DeNardo:1996kp} 
for gauge fields calculated by introducing the conical singularity.  

If we use the hyperbolic heat kernels  
 \cite{CH1994} \cite{GN} \cite{Giombi:2008vd} to consider the 4D conformally 
coupled scalar field, the log-divergent term in the R\'enyi 
entropy can be obtained directly. We obtain
\bea
S^{(s=0)}_{\rm{q} ,\text{ln}}=-{1+\rm{q}+\rm{q}^2+ \rm{q}^3\over {360 \rm{q}^3}}\ln ({{\rm{R}}\over \delta }) \ .
\eea 
Taking $\rm{q}=1$ gives 
\bea
S^{(s={0})}_{{\rm{EE}},\text{ln}}=-{1\over 90}~\ln ({{\rm{R}}\over \delta}) \ ,
\eea 
which matches exactly with the expected type-A anomaly
prediction. On the other hand, there is no mismatch problem for fermions. 
The heat kernel and related algebra can be found in literature 
(for example, see the appendix in \cite{Hung:2014npa}).
For useful reference, we list the corresponding result of the 4D Dirac fermion,
\bea
S^{(s={1\over 2})}_{\rm{q},\text{ln}}=-{7+7\rm{q}+37 \rm{q}^2+ 37 \rm{q}^3\over {720 \rm{q}^3}}\ln ({{\rm{R}}\over \delta }) \ .
\eea 
Taking $\rm{q}=1$, it gives the expected conformal anomaly prediction 
\bea
S^{(s={1\over 2})}_{{\rm{EE}},\text{ln}}=-{11\over 90}~\ln ({{\rm{R}}\over \delta}) \ .
\eea 
In short, the field-theory calculation of the log-divergent terms 
of the conformally coupled scalar field and 
massless fermion on ${\rm{S}}^1 \times {\rm {H}}^{d-1}$ match directly with 
the conformal anomaly prediction, without any edge correction needed. 
This is consistent with the fact that the boundary modes contribute 
in the gauge-field case due to the existence of the gauge symmetry.
\\

\noindent
{\it{\bf{6: Concluding Remarks}}}

Let us make a remark on the contribution of the edge modes to the entropy 
calculation in the static patch of de Sitter space. 
Starting again with the flat-space metric \eqref{flat}, one uses the  
coordinate transformation 
 \bea
t&=&{\rm{R}}\,\frac{\cos\theta\,\sinh({\tau\over {\rm{R}}})}{1+\cos\theta\,\cosh({\tau\over {\rm{R}}})}\ , \\ 
r&=&{\rm{R}}\,\frac{\sin\theta}{1+\cos\theta\,\cosh({\tau\over {\rm{R}}})}\ ,
 \eea 
to obtain 
 \bea
\label{ds}
{\rm{d}}s^2=\Omega^2\, \left[-\cos^2\!\theta\, {\rm{d}}\tau^2+{\rm{R}}^2\left({\rm{d}}\theta^2+\sin^2\theta\,{\rm{d}}\Omega^2_{d-2}\right)\right] \ ,
 \eea 
where $\Omega=\Big(1+\cos\theta\,\cosh({\tau\over {\rm{R}}})\Big)^{-1}$ is the 
conformal factor to be eliminated. The remaining metric
is the static patch of de Sitter space with the scale ${\rm{R}}$.  
The important limits are
\bea
\tau=\pm \infty  \rightarrow&&(t,r)=(\pm {\rm{R}},0) \ ,\\
\theta=\frac{\pi}{2} \rightarrow &&(t,r)=(0,{\rm{R}})\ .
\eea
We see again that the new coordinates cover the causal development
$\cal D$ of the ball $r \leq {\rm{R}}$ at $t=0$. As shown in \cite{Casini:2011kv}, similar 
to the hyperbolic space, the modular transformation inside $\cal D$ again corresponds 
to the time translation in de Sitter space after the 
conformal mapping. The state in de Sitter geometry becomes thermal at $T=1/(2\pi {\rm {R}})$. 
The thermal entropy in de Sitter space then can be also identified as the entanglement 
entropy in flat spacetime with a codimension-2 spherical entangling boundary.

In \cite{Dowker:2010bu}, the thermal entropy in $4 \rm{D}$ de Sitter space is calculated. 
Interestingly, the same mismatch \eqref{mis} is found in the log-divergent term. 
The volume of de Sitter space is finite, so in this case there is no need to 
introduce an IR cut-off. The log divergence comes from the UV divergence of  
the partition function. In calculating the partition function of gauge fields, the gauge 
fixing process again results in the gauge redundancy. 
Notice that in this case, one identifies the entangling boundary 
as the cosmological horizon at the boundary of the static patch at $\theta=\pi/2$.  
From the metric \eqref{ds}, we see that the boundary defined by the limit $\theta \to \pi/2$ 
is again effectively  
a 2-sphere since the prefactor of the time direction shrinks to zero in this limit. 
The boundary redundancy modes 
effectively satisfy (in the spirit of Kaluza-Klein dimensional compactification)
again \eqref{key}, which gives the same edge correction \eqref{13} that resolves the  
mismatch found in the de Sitter space calculation as well.

It would be of great interest to better understand the edge 
modes and explore its potential applications. It has been suggested  \cite{Buividovich:2008gq} 
(see also \cite{Donnelly:2008vx} \cite{Casini:2013rba})
that one might modify the Hilbert space decomposition as
\bea
\cal{H}=\cal{H_{\rm{A}}}\otimes \cal{H_{\rm{B}}} \otimes \cal{H_{\partial {\rm{A}}}} \ ,
\eea
where $\cal{H_{\partial {\rm{A}}}}$ denotes a boundary Hilbert space, to have a
special treatment of boundary in calculating the entanglement entropy of gauge fields.
Let us make an initial attempt to relate this idea to the approach considered here.
If one wants to derive the edge contribution \eqref{13} starting from a classical surface action,
an immediate issue is that a surface action will cause trouble 
having a well-defined variational principle. If we introduce a boundary Hilbert space 
separately, we might consider a surface action subjected to the path integral quantization in this
separated Hilbert space. Then, to incorporate the contribution from the 
edge, we might identify the edge partition function as
\bea
\label{surface}
Z_{\rm{Edge}}= Z^{1\over 2}_{\rm{Ghosts}}~;~
Z_{\rm{Ghosts}}= \int {\cal D}\bar b {\cal D} b e^{ -\int_{S^2} (\bar b \Box b)}\ .~~
\eea 
(Since the edge part does not have
any gauge field, we simply define $\delta_{\eps} { b}_{|\rm{S}^2}=\delta_{\eps} \bar { b}_{|\rm{S}^2} =0$ 
so that the surface action remains BRST invariant.)
Notice that because of the intrinsic asymmetric treatment on ghost fields 
$b$ and $\bar { b}$ in the BRST symmetry, the gauge redundancy (and
the resulting edge correction on the entropy) is solely determined by ${ b}$ in this framework,
so we adopt $Z^{1\over 2}_{\rm{Ghosts}}$ as the correct counting.  

Finally, we would like to point out that the edge modes are introduced in 
the context of black hole entropy in \cite{Keeler:2014bra}, where the 
authors consider BPS black holes with $\rm{AdS}_2 \times {\rm{S}}^2$ near horizon 
geometry. They argue that gauge symmetries give rise to physical modes 
that localize on the boundary of $\rm{AdS}_2$ and contribute to the black-hole 
entropy.\footnote{I thank Finn Larsen for explaining some concepts in \cite{Keeler:2014bra}.} 
The boundary modes in their context come also from supersymmetry and 
diffeomorphism invariance. It is shown that the boundary modes are needed 
to obtained the expected log corrections to the black-hole entropy. While 
it remains to further explore the deeper relation between the black-hole entropy 
and the entanglement entropy, it would be nice to see that the edge modes 
are essential in the both contexts. It would be interesting to calculate the entanglement 
entropy of gravitons or other types of gauge fields using the approach considered 
in this paper, without introducing the conical singularity, and see if the 
corresponding entanglement edge modes due to gauge redundancies 
are needed to explain possible discrepancies.
\\
\\

\noindent
{\it{\bf{Acknowledgments}}}

I would like to thank Dmitri Fursaev, Finn Larsen, Sergey Solodukhin and 
especially Christopher Herzog for discussions and comments. 
I also thank Christopher Akers and Aitor Lewkowycz for dicussions on boundary terms.
This work was supported in part by the National Science
Foundation under Grant No. PHY13-16617.
\\

$Note~ added.$--- Recently, a paper \cite{Donnelly:2014fua} 
appeared that considered the entanglement entropy of gauge 
fields using an approach in which the conical singularity was introduced.
The authors of that paper related the 4D mismatch result with the entangling 
boundary independently.

\end{document}